


\magnification\magstep1
\parskip=\medskipamount
\hsize=6.0 truein
\vsize=8.2 truein
\hoffset=.2 truein
\voffset=0.4truein
\baselineskip=14pt
\tolerance 500


\font\titlefont=cmbx12
 at 10 truept
\font\authorfont=cmcsc10
\font\addressfont=cmsl10 at 10 truept
\font\smallbf=cmbx10 at 10 truept


\outer\def\beginsection#1\par{\vskip0pt plus.2\vsize\penalty-150
\vskip0pt plus-.2\vsize\vskip1.2truecm\vskip\parskip
\message{#1}\leftline{\bf#1}\nobreak\smallskip\noindent}


\newdimen\itemindent \itemindent=13pt
\def\textindent#1{\parindent=\itemindent\let\par=\resetpar%
\indent\llap{#1\enspace}\ignorespaces}

\let\oldpar=\par
\def\resetpar{\oldpar\parindent=0pt\let\par=\oldpar}

\font\ninerm=cmr9 \font\ninesy=cmsy9
\font\eightrm=cmr8 \font\sixrm=cmr6
\font\eighti=cmmi8 \font\sixi=cmmi6
\font\eightsy=cmsy8 \font\sixsy=cmsy6
\font\eightbf=cmbx8 \font\sixbf=cmbx6
\font\eightit=cmti8
\def\eightpoint{\def\rm{\fam0\eightrm}
  \textfont0=\eightrm \scriptfont0=\sixrm \scriptscriptfont0=\fiverm
  \textfont1=\eighti  \scriptfont1=\sixi  \scriptscriptfont1=\fivei
  \textfont2=\eightsy \scriptfont2=\sixsy \scriptscriptfont2=\fivesy
  \textfont3=\tenex   \scriptfont3=\tenex \scriptscriptfont3=\tenex
  \textfont\itfam=\eightit  \def\it{\fam\itfam\eightit}%
  \textfont\bffam=\eightbf  \scriptfont\bffam=\sixbf
  \scriptscriptfont\bffam=\fivebf  \def\bf{\fam\bffam\eightbf}%
  \normalbaselineskip=9pt
  \setbox\strutbox=\hbox{\vrule height7pt depth2pt width0pt}%
  \let\big=\eightbig \normalbaselines\rm}
\catcode`@=11 %
\def\eightbig#1{{\hbox{$\textfont0=\ninerm\textfont2=\ninesy
  \left#1\vbox to6.5pt{}\right.\n@space$}}}
\def\vfootnote#1{\insert\footins\bgroup\eightpoint
  \interlinepenalty=\interfootnotelinepenalty
  \splittopskip=\ht\strutbox %
  \splitmaxdepth=\dp\strutbox %
  \leftskip=0pt \rightskip=0pt \spaceskip=0pt \xspaceskip=0pt
  \textindent{#1}\footstrut\futurelet\next\fo@t}
\catcode`@=12 %


\def\mapright#1{\smash{
  \mathop{\longrightarrow}\limits^{#1}}}
\def\mapdown#1{\Big\downarrow
  \rlap{$\vcenter{\hbox{$\scriptstyle#1$}}$}}
\def\diag{
  \def\normalbaselines{\baselineskip2pt \lineskip3pt
    \lineskiplimit3pt}
  \matrix}


\def\Ds{/\!\!\!\! D}
\def\Dts{/\!\!\!\! {\hat D}}
\def\Gs{{\cal G}^{2n}}
\def\Gst{{\cal G}^{2n+1}}
\def\As{{\cal A}^{2n}}
\def\Ast{{\cal A}^{2n+1}}
\def\Qs{{\cal Q}^{2n}}
\def\Qst{{\cal Q}^{2n+1}}

\def\shalf{\hbox{${\textstyle{1\over 2}}$}}

\tolerance=500


\rightline{Freiburg, THEP-94/12}
\rightline{hep-th/9405178}
\bigskip
{\baselineskip=24 truept
\titlefont
\centerline{INDUCED CONNECTIONS IN FIELD THEORY:}
\centerline{THE ODD-DIMENSIONAL YANG-MILLS CASE}
}

\vskip 1.1 truecm plus .3 truecm minus .2 truecm

\centerline{\authorfont Domenico Giulini\footnote*{
e-mail: giulini@sun1.ruf.uni-freiburg.de}}
\vskip 2 truemm
{\baselineskip=12truept
\addressfont
\centerline{Fakult\"at f\"ur Physik,
Universit\"at Freiburg}
\centerline{Hermann-Herder Strasse 3, D-79104 Freiburg, Germany}
}
\vskip 1.5 truecm plus .3 truecm minus .2 truecm

\centerline{\smallbf Abstract}
\vskip 1 truemm
{\baselineskip=12truept
\leftskip=3truepc
\rightskip=3truepc
\parindent=0pt

{\eightpoint
We consider $SU(N)$ Yang-Mills theories in $(2n+1)$-dimensional Euclidean
spacetime, where $N\geq n+1$, coupled to an even flavour number of Dirac
fermions. After integration over the fermionic degrees of freedom
the wave functional for the gauge field inherits a non-trivial
$U(1)$-connection which we compute in the limit of infinite fermion
mass. Its Chern-class turns out to be just half the flavour number
so that the wave functional now becomes a section in a non-trivial
complex line bundle. The topological origin of this phenomenon is
explained in both the Lagrangean and the Hamiltonian picture.
\par}}

\beginsection Introduction

Induced connections can arise in any theory whose classical configuration
space displays a certain topological richness. If these connections
have non-trivial curvature -- the case we are interested in --
the precise condition is that the second cohomology $H^2(Q,Z)$
of the classical configuration space $Q$ must have a non-trivial free
part (i.e. factors of $Z$). Rather than outlining the general theory
(which is basically the classification of $U(1)$-principal bundles over
$Q$, lucidly explained e.g. in Ref. 1) we try to develop some feeling
for the underlying mechanism and assumptions, by first discussing a simple
finite dimensional toy model which mimics exactly the essential features
without the analytic complications. Similar toy models have been used
extensively throughout the literature in explaining the geometric and
topological origin of anomalies and Berry-phases. But eventually we are
interested in field theory. There we wish to understand in detail how
the mere possibility of induced connections -- established by pure
topological arguments -- is actualized by concretely given dynamical
laws. The purpose of this article is to present such an example in
which this process can be studied in detail. Similar to so many
contributions to the understanding of anomalies, it will once more
show a deep link between topology, geometry and dynamics.

\beginsection Section 1: A Toy Model

Consider the 2-dimensional Hilbert space, $C^2$, and a Hamiltonian
$$
H=x\cdot\tau\,,
\eqno{(1.2)}
$$
parameterized by the 2-sphere $S^2=\{x\in R^3\,/\,\|x\|=1\}$.
$\tau=(\tau_1,\tau_2,\tau_3)$ are the Pauli matrices. We think of $S^2$
as the position-space of some particle. If we calculate the
eigenvectors of $H$, we find that there is no global phase choice to make them
well defined over the whole of the 2-sphere. To circumvent this, we think of
$S^2$ as the quotient of $S^3\cong SU(2)$ via the action of $U(1)_R$, the
group of right translations generated by ${i\over 2}\tau_3$. Let $g$ be a
general element of $SU(2)$. The quotient map (the Hopf map) is then given by
$$
\eqalignno{
\pi:\; S^3 & \rightarrow S^2 \cr
       g\, & \mapsto g\tau_3 g^{-1}\,=x\cdot\tau &(1.2)\cr}
$$
such that the Hamiltonian is now given by $g\tau_3 g^{-1}$, which is clearly
invariant under $U(1)_R$. Its eigenvalues
are $\pm 1$ with eigenvectors,  $|\pm\,,g \rangle$, given by
$$
|\pm\,,g \rangle \,=\,g\cdot e_{\pm}\,,\quad e_+=\pmatrix{1\cr 0\cr}\quad
e_-=\pmatrix{0\cr 1\cr}\,.
\eqno{(1.3)}
$$
That is, the positive eigenvector is given by the first column of the
matrix $g$, the negative one by the second. We write this as
$$\eqalign{
& |+\,,g\rangle =\{g_{i1}\}\quad,\quad |-\,,g\rangle =\{g_{i2}\}\,,
 \cr
 \hbox{such that}\quad
& |\pm\,, g\exp{(i\shalf}\alpha\tau_3)\rangle
  =\exp(\pm{i\shalf}\alpha)|\pm\,g\rangle\,,
 \cr}
\eqno{(1.4)}
$$
where the last equation specifies the representations, $\rho_{\pm}$, of
$U(1)_R$.
Infinitesimally, the $g$-dependence of $|g,\pm\rangle$ can be written as
$$
dg_{lk}=\Theta^n_k\,g_{ln}
\eqno{(1.5)}
$$
where $\Theta$ is the matrix of left invariant 1-forms, $\{\sigma^i\}$, on
$SU(2)$:
$$
\Theta=g^{-1}\,dg={i\shalf}\,\sigma\cdot\tau \,.
\eqno{(1.6)}
$$
The eigenvectors, $|\,+\,\rangle$ and $|\,-\,\rangle$, are equivariant
functions on $S^3$ into the Hilbert-eigenspaces ${\cal H}_{\pm}$
(here 1-dimensional) carrying the representation, $\rho_{+}$, and its
complex conjugate, $\rho_{-}$, respectively. This can be expressed by
the commutative diagram:
$$
\diag{
S^3&\mapright{{}_{|\pm\rangle}}&{\cal H}_{\pm}\cr
\mapdown{{}_{U(1)_R}}&&\mapdown{\rho_{\pm}}\cr
S^3&\mapright{{}_{|\pm\rangle}}&{\cal H}_{\pm}\cr}
\eqno{(1.7)}
$$

If we now want to quantize the $S^2$-degree of freedom as well (i.e.
the particle motion) we can instead use $S^3$ as enlarged configuration
space with one redundant (gauge) degree of freedom. We then have a wave
function
$$
\psi:\;S^3\longrightarrow {\cal H}_+\oplus{\cal H}_-
\eqno{(1.8)}
$$
which, in the adiabatic approximation, (i.e., under the hypothesis of a slowly
moving particle) can be split into $\psi_{\pm}$, each mapping into the
1- dimensional ${\cal H}_{\pm}$ only. The redundant degree of freedom, which
is not to be quantized, is then taken care of by imposing the
``Gauss constraint'' which just expresses the adiabaticity condition
in the requirement that $\psi_{\pm}$ be an equivariant, ${\cal H}_{\pm}$-
valued function on $S^3$, or equivalently, a section in a nontrivial line
bundle over $S^2$, associated to the Hopf bundle (1.2) in the
representation $\rho_{\pm}$.  According to (1.7) the Gauss constraint
thus reads
$$
X_3\psi_{\pm}=\pm{i\shalf}\psi_{\pm}\,,
\eqno{(1.9)}
$$
with $X_3$ being the left invariant vector field dual to $\sigma_3$.
The last step is now to implement (1.9) dynamically, i.e. to find a
Lagrangean for the particle on $S^3$ which has $p_3=\pm{1\over2}$ as a
constraint. To do this, we write down the standard line
element on $S^2$ in terms of the 1-forms $\{\sigma^i\}$
$$
ds^2=\shalf\left[\,\sigma_1^2\,+\,\sigma_2^2\,\right]\,,
\eqno{(1.10)}
$$
which is $SU(2)_L\times U(1)_R$ invariant. The unique term with the same
symmetries that gives the desired constraint is $\pm{1\over 2}\sigma_3$.
The ``effective'' Lagrangean can then be locally projected onto $S^2$ and
reads in the coordinate system that covers the 2-sphere except at the north
pole
$$
L=\shalf m|\dot x|^2\,\pm\,\shalf(\,1+\cos\theta)\dot\varphi\,.
\eqno{(1.11)}
$$
But this is just the Lagrangean of an electrically charged particle of unit
charge in the background of a magnetic monopole of strength $g={1\over 2}$.
Recall that the connection (gauge potential) has been deduced under the
explicit assumption of slow motion (hypothesis of adiabaticity). It might
well be called the {\sl adiabatic connection}, and its holonomies are just
the celebrated Berry-Phases. For the field theoretic model of the next
section it will be useful to have the following correspondences in mind:
$$\eqalign{
{\cal H}=C^2 & \longleftrightarrow \hbox{fermionic Hilbert space}      \cr
S^3          & \longleftrightarrow \hbox{space of gauge potentials:}\,
                                        {\cal A}                       \cr
S^2          & \longleftrightarrow \hbox{space of gauge orbits:}\,
                                        {\cal A}/{\cal G}={\cal Q}\,.  \cr}
\eqno{(1.12)}
$$
In this model massive fermions will induce a connection
on the effective gauge theory which will be the adiabatic connection in
the limit of infinite fermion mass. In the toy model we had
$H^2(S^2,Z)=Z$ and the ``magnetic field of the monopole''
represented a non-trivial class therein (using DeRahm's construction).
The same picture arises in the infinite dimensional model.
(See Ref. 2 for a general discussion of how this
topological class is also generally responsible for a specific type of
anomalies.)

\beginsection Section 2: The 2n+1-Dimensional Yang-Mills Case

In this section we consider 2n+1 dimensional Euclidean Yang Mills  fields
coupled to Dirac fermions. The gauge group is taken to be $G=SU(N)$ where
$N\geq 2$. Since Euclidean space is contractible, the bundle is necessarily
trivial. Elements of the group of gauge-transformations are given by
ordinary $SU(N)$-valued functions.
By standard arguments gauge transformations are restricted to be the
identity at infinity. This holds for space-time gauge transformations,
as well as just spatial ones.
In the Hamiltonian formulation we go into the $A_0=0$ gauge and consider
instead a $G$-bundle over the $t=$const. slices (called $\Sigma$).
In both cases, the group of gauge transformations may be identified with
the function space of the form
$$
\left( R^N,\infty\right)\mathop{\longrightarrow}^{g} \left(G,e\right)\,,
\eqno{(2.1)}
$$
where $N$ is either $2n+1$ or $2n$ and $e$ is the identity in $G$. In the
first case we call it $\Gst$, in the second $\Gs$. The symbol $\infty$
denotes the infinity in Euclidean space and the gauge transformations
map the point at infinity to the identity $e$. The spaces of gauge
potentials are called $\Ast$ and $\As$ respectively. For statements
which are true in either case, we shall omit the superscripts.

${\cal G}$ acts on ${\cal A}$ via
$$
g\times A\longmapsto g^{-1}(A+d)g^1\,.
\eqno{(2.2)}
$$
If $A\in {\cal A}$ is fixed by the action of ${\cal G}$, it obeys $Dg=0$,
where $D$ is the exterior covariant derivative. With the imposed boundary
conditions it is easy\footnote{${}^{a}$}{The best way to see this is not
to look at
the gauge potentials but at the connection 1-forms on the principal bundle,
the former being pull-backs of the latter by some local sections. On the
total space $P$, gauge transformations are given by bundle automorphisms
projecting to the identity, or, equivalently, by
$G$ valued functions on $P$ which are Ad-equivariant under the right action of
$G$ on $P$. Covariant constancy now means that the differential of
this matrix valued function is zero if restricted to horizontal subspaces. The
boundary conditions at infinity then force it to be the unit matrix.} to see
that this implies $g\equiv e$. ${\cal G}$ therefore acts freely and
${\cal A}$ can be given the structure of a principal fibre bundle
(see Ref. 3)
$$\diag{
{\cal G}&\mapright{}&{\cal A}\cr
&&\mapdown {\tau}\cr
&&{\cal Q}\cr}\eqno(2.3)
$$
where we have in fact two spaces, $\Qst$ and $\Qs$. Only the latter acts
as configuration space in the canonical formulation and we will simply call it
the configuration space.
Since ${\cal A}$ is an affine space, we have from the associated exact
homotopy sequence
$$\eqalignno{
&\pi_k(\Qs)=\,\pi_{k-1}(\Gs)=\pi_{2n+k-1}(G)     &(2.4)\cr
&\pi_k(\Qst)=\,\pi_{k-1}(\Gst)=\pi_{2n+k}(G)\,.  &(2.5)\cr}
$$
In particular we have, now specializing\footnote{${}^{b}$}{
Bott-periodicity implies
$\pi_{2n}(SU(N))=0$ and $\pi_{2n+1}(SU(N))=Z$ for $N\geq n+1$.}
$N$ to $N\geq n+1$,
$$\eqalignno{
&\pi_0(\Gs)=\pi_1(\Qs)=1= \pi_1(\Gst)=\pi_2(\Qst)    &(2.6)\cr
&\pi_0(\Gst)=\pi_1(\Qst)=Z= \pi_1(\Gs)=\pi_2(\Qs)\,, &(2.7)\cr}
$$
which also implies $H^2(\Qs)=Z$ and hence the possibility of monopoles
in the configuration space $\Qs$. It is the purpose of the rest of this
paper to demonstrate that the interaction with matter (here massive
Dirac fermions) causes the wave function for the Yang Mills field to
actualize this topological possibility. In Ref. 4 the possibility of
induced connections has been anticipated and their consequences for the
equal-time commutation relations discussed.
In our derivation we follow the spirit of Ref. 2 and make
use of both, the Lagrangean formulation, where gauge fields are
defined over space time $M$, and the Hamiltonian formulation, where via
the gauge condition $A_0=0$ one has a gauge theory over the spatial
sections $\Sigma$. Note that a gauge transformation in $\Ast$ is given
by a function $g:\,[0,1]\times S^{2n} \rightarrow G$, such that
$g_1=g_0\equiv e$ and $g_t(\infty)=e\,\forall t$, which at the same time
defines an element of $\pi_1(\Gs)$. Therefore, given a
non-closed path  in $\Ast$ which connects two different components of
$\Gst$ in such a way  that it projects to a  loop in $\Qst$ which
generates $\pi_1(\Qst)$, one  has at the same time found a generator of
$\pi_1(\Gs)$. In $\As$ this generator is the boundary of a 2-disk whose
image (under the quotient map $\As\rightarrow \Qs$) is a
non-contractible 2-sphere generating $\pi_2(\Qs)$.

Finally, let us note that the bundle
(2.3) with group $\Gs$ total space $\As$ and base $\Qs$ can be given a
natural connection once the metric on the spatial slices has been specified.
Tangent vectors in $A\in \As$ are Lie algebra-valued one forms which under
gauge transformations transform with the inverse\footnote{${}^{c}$}{
In our convention
gauge transformations are right actions.} adjoint representation. We call
this space $\Lambda^1(LieG)$. Let $T_A(\As)=V_A\oplus H_A$ be an orthogonal
decomposition of the tangent space at $A$. $V_A$ is
the vertical space spanned by vectors of the form $X^{\omega}_A
=D_A\omega$, where $D_A$ is the covariant derivative at $A$ and $\omega$ is an
element of $\Lambda^0(Lie G)$, the Lie algebra of $\Gs$. $H_A$ is by
definition the orthogonal complement of $V_A$ using the metric $r_A$,
defined by ($*$ is the Hodge-duality map)
$$
r_A(Y_A,Z_A):=\int_{\Sigma} {\rm tr}\left(Y_A\wedge * Z_A\right)\,.
\eqno(2.8)
$$
$r$ is invariant under the action of $\Gs$ and hence $H_A$ defines a
connection. Locally $H_A$ can be expressed as the kernel of the operator
$D^{\dagger}_A$, which is the adjoint of $D_A$ with respect to $r$.
The connection 1-form can then be written as (see Ref. 5)
$$
C_A:=G_A\circ D^{\dagger}_A\;,\;
\hbox{where}\;G_A=(D^{\dagger}_A \ D_A)^{-1}\,.
\eqno(2.9)
$$
It annihilates elements of $H_A$, transforms in the appropriate form under
gauge
transformations in $\Gs$, and, when acting upon vertical vector fields
$X^{\omega}_A$, one has
$$
C_A(X^{\omega}_A)=\omega \in \Lambda^0(LieG)
\eqno{(2.10)}
$$
as required for connections. The metric (2.8) and the connection (2.9) have
already been used in attempts to geometrically understand anomalies and
also to formulate a Riemannian geometry of $\Qs$ (Refs. 5,6,7).
\medskip

The Euclidean action for Yang-Mills coupled to Dirac fermions is given
by\footnote{${}^{d}$}{We use the convention
$\{\gamma^{\mu},\gamma^{\nu}\}= -2\delta^{\mu\nu}$. The $\gamma^{\mu}$'s
are thus anti-Hermitean $2^n\times 2^n$-matrices.}
$$
S_E=\int_M d^{2n+1}x\,\left[{1\over4}tr(F_{\mu\nu}F^{\mu\nu})
    -\bar{\psi}(i\Ds\,-\,m)\psi\right]\,,
\eqno(2.11)
$$
where the fermions carry in addition to their spin and Lie group index
also a flavour index of dimension $f$.
The path integral over the fermions defines an effective action for
the gauge field $A$ by
$$
\int dA\;\int d\bar{\psi}d\psi \,{\rm e}^{-S_E[A,\bar{\psi},\psi]}
=:\int dA\,{\rm e}^{-S_{\rm eff}[A]}\,.
\eqno{(2.12)}
$$

We wish to determine $W[A]=\ln Z[A]$, where
$$
Z[A] :=\int d\bar{\psi}d\psi\,{\rm e}^{\int \bar{\psi}(i\Ds-m)\psi }\,.
\eqno(2.13)
$$
We expand the connection in terms of Hermitean basis matrices
$\{T_1,\dots ,T_k\}$, $k=dim SU(N)$, so that $A=iT_pA^p_{\mu}\,dx^{\mu}$,
and define the current by
$$
I^{\mu}_p[A]={\delta\over \delta A^p_{\mu}}W[A]\,.
\eqno(2.14)
$$
By construction only the exponential of $W[A]$ is expected to give a
well defined function on $\Qst$. A method to obtain local expressions
for $W[A]$ is to calculate the one-form $\delta W[A]$ at a preferred
point $A$ and then integrate this expression within a simply connected
neighbourhood of $A$. We shall follow this strategy in the appendix.
The obstruction to extend this to a globally
defined function $W[A]$ is given by the cohomology class in $H^1(\Qst)$
generated by the one form $\delta W$.

Using known techniques, we calculate $I^{\mu}_A$ in a ${1\over m}$ -
expansion. The zeroth order term (i.e. the $m\rightarrow\infty$ limit)
then gives us the adiabatic connection. The calculation, which we defer
to the appendix, yields for the zeroth order term (compare formula (A.14)
from the appendix)
$$
I^0_p[A]=-f{i\over 2}{1\over n!}
\left({i\over 2\pi}\right)^n\,{1\over 2^n}
\varepsilon^{0i_1\dots i_{2n}}\,{\rm tr}
\left(T_p\,F_{i_1i_2}\dots F_{i_{2n-1}i_{2n}}\right)\,.
\eqno(2.15)
$$
Here $T$ is a basis element of $LieG$ and the trace is taken over
the Lie algebra indices. It follows that $I^0_A$ transforms with the inverse
co-adjoint representation under gauge transformations which it should do being
an element of the dual of the Lie algebra
of $\Gs$. We shall use this fact later in the canonical picture.
Here we shall follow the original plan and insert the result in (2.14) to
obtain
$$
\delta\ W[A]=2\pi i\,{f\over 2}{1\over n!}
\left({i\over 2\pi}\right)^{n+1}\,\int_M
{\rm tr}\left(\delta A\; F^n [A]\right)\,.
\eqno(2.16)
$$
According to the discussion following (2.7) we now want to integrate
(2.16) along a generator of $\pi_1(\Qst)$, i.e. along a curve that starts and
ends in different components of $\Gst$. For this we can take any starting point
$A$ in $\Ast$ since the result does not depend on it. Following the standard
notations (see e.g. Ref. 8), we call
$$\eqalignno{
&\omega^1_{2n+1}[A]   :=(n+1)\int_M\,{\rm tr}\,(\delta A\,F^n [A]) &(2.17)\cr
&\Omega^1_{2n+1}(0,A) := \int_{\gamma(0,A)}\omega^1_{2n+1} \,,     &(2.18)\cr}
$$
where in the first equation we have defined a gauge invariant closed
1-form in $\Ast$, which defines therefore a closed 1-form in $\Qst$, and
in the second equation we integrated this 1-form over a straight path from 0
to $A$ which we denoted by $\gamma(0,A)$. $\Omega(0,A)$ is also known as the
first Chern-Simons form.

We now integrate the 1-form $\omega^1_{2n+1}$ along the edges of two different
triangles in
$\Ast$. The first one has vertices $(0,A,A^g)$, the second $(0, g^{-1}dg,
A^g)$.
Since in $\Ast$ closed forms are necessarily exact, the two integrals are zero.
We thus arrive at the two relations
$$
\eqalignno{
\Omega^1_{2n+1}(A,A^g)      &=
\Omega^1_{2n+1}(0,A^g)-\Omega^1_{2n+1}(0,A)             &(2.19) \cr
\Omega^1_{2n+1}(0,g^{-1}dg) &=
\Omega^1_{2n+1}(0,A^g)-\Omega^1_{2n+1}(g^{-1}dg,A^g)\,. &(2.20) \cr}
$$
Invariance of $\Omega^1_{2n+1}$ under simultaneous gauge transformations
in both arguments implies equality  of the last terms in each line, and
hence equality of the expressions on the left sides of (2.19) and (2.20).
Since $\Omega^1_{2n+1}(A,A^g)$ is the line
integral of $\omega^1_{2n+1}$ from $A$ to $A^g$, it also represents the loop
integral of the projected 1- form on $\Qst$. This 1-form  generates
$H^1(\Qst)$ if its integral along a generator of $\pi_1(\Qst)$ gives the result
1.
For this,  $g$ has to be a gauge transformation of unit winding number.
The desired expression for this
integral is now seen to be given by the integral along the straight path
between
0 and $g^{-1}dg$ which is independent of $A$ as required. Elementary
integration yields
$$
\int_{\gamma(0,g^{-1}dg)}\omega^1_{2n+1}=(-1)^n{n!(n+1)!\over (2n+1)!}\int_M
{\rm tr}(g^{-1}dg)^{2n+1}\,.
\eqno(2.21)
$$
On the other hand, the integer-valued winding number $w(g)$ of $g$ is
given by the expression (see Ref. 9)
$$
w[g]=\left({i\over 2\pi}\right)^{n+1}{n!\over (2n+1)!}\int_{M}{\rm
tr}(g^{-1}dg)^{2n+1}
\eqno(2.22)
$$
so that we finally arrive at the following expression for the line integral of
$W[A]$ along a generator of $w(g)\cdot Z$
$$
\oint\delta W[A]=(-1)^n2\pi i\,{f\over 2} w[g]\,.
\eqno(2.23)
$$
We notice that $\exp(W[A])$ will only be a well defined
function on $\Qst$ if $f$ is even. In this case, $\delta W[A]$ represents the
integer class $f/2$ in $H^1(\Qst)=Z$.

Let us now turn to the Hamiltonian picture. For this, we go back to
(2.15). There we expect to find encoded the same information in $H^2(\Qs)$
represented by some curvature 2-form. The infinitesimal holonomy enters the
physical picture by anomalous commutators (Schwinger terms). Let us try to
explain this in the geometric picture developed so far.

If the action (2.11) is put into canonical form, the first class
constraint associated with the gauge freedom in $\As$ appears as
Gauss' law
$$
D_k\pi^k=\bar{\psi}\gamma^0T\psi\,,\qquad\pi^k=F^{0k}\,.
\eqno{(2.24)}
$$
In an effective theory for the gauge field the right hand side is replaced by
its expectation value $I^0_A$. Quantizing the gauge field in the Schr\"odinger
picture involves the constraint (here the dot $\cdot$ represents summation and
integration)
$$\eqalignno{
& \left[X^{\omega}\,-i\omega\cdot I^0\right]\psi[A]=0\,,
& (2.25)\cr
\hbox{where}\quad
& X^{\omega}[A]=D_A\omega\cdot{\delta\over \delta A}
&(2.26)\cr}
$$
are the fundamental vector fields on the principal bundle $\As$. It is
easy to verify that the map $\omega\mapsto X^{\omega}$ furnishes a
homomorphism from the Lie algebra of $\Gs$ into the Lie algebra of vector
fields on $\As$. With the aid of the connection $C_A$ from equation (2.9)
and the charge density $I^0_A$ we can form the $\Gs$-invariant 1-form on
$\As$
$$\eqalignno{
& \Omega=I^0\cdot C\,,  &(2.27)\cr
\hbox{satisfying}\quad
& \Omega_A(X^{\omega}_A)=\omega\cdot I^0_A\,, &(2.28)\cr}
$$
so that (2.25) can now be written in the form of a parallel
transportation law along all vertical directions with respect
to the $U(1)$-connection $-i\Omega$, where, as usual, we separated an
imaginary unit to have $\Omega$ real valued. The corresponding covariant
derivative operator is called $\nabla$:
$$
\left[X^{\omega}\,-i\Omega(X^{\omega})\right]\psi[A]
 =:\nabla_{X^{\omega}}\psi[A]=0 \,.
\eqno(2.29)
$$

In quantum field theory, an {\it anomalous commutator} is defined as
the deficiency term that prevents the mapping
$\omega\mapsto\nabla_{X^{\omega}}$ from the Lie algebra of $\Gs$ to the
commutator algebra of linear operators on quantum states from being a
homomorphism of Lie algebras:
$$
\left[\nabla_{X^{\omega}},\nabla_{X^{\eta}}\right]_{\rm anom}:=
\left[\nabla_{X^{\omega}},\nabla_{X^{\eta}}\right]\,-\,\nabla_{[X^{\omega},
X^{\eta}]}\,,
\eqno(2.30)
$$
where we used that $X^{[\omega,\eta]}=[X^{\omega},X^{\eta}]$.
But the right hand side is just the curvature two-form --denoted by $K$--
for the $U(1)$-connection $\Omega$, evaluated on the fundamental
vector fields $X^{\omega}$ and $X^{\eta}$. It satisfies
$$
K(X,Y)=d\Omega(X,Y):=X[\Omega(Y)]-Y[\Omega(X)]-\Omega([X,Y])\,.
\eqno(2.31)
$$
Using (2.15), we write the connection as follows:
$$
\Omega_A=-{f\over 2}2\pi\,{1\over n!}\left({i\over 2\pi}\right)^{n+1}
\int_{\Sigma} {\rm tr}\,\left(C_A\ F^n[A]\right)\,.
\eqno(2.32)
$$
Integrating $K$ over a 2-sphere in $\Qs$ can be done by integrating our
expression
for $K$ over a disc in $\As$ with boundary in a fibre $\Gs$, or, equivalently,
by  integrating $\Omega$ over the boundary circle. To do this, let $g(t)$ be
a loop in $\Gs$ and $\gamma(t):=A^{g(t)}$ the associated loop through $A$ in
$\As$. The generating vector field is given by
$$
\gamma_*{d\over dt}=D_{\gamma(t)}\left(g^{-1}{d\over dt}g\right)
\eqno{(2.33)}
$$
We now integrate $\Omega$ along this vector field and obtain:
$$
\oint_t\gamma^*\Omega =-2\pi{f\over 2}\,{1\over n!}
\left({i\over2\pi}\right)^{n+1}
\int_{S^1}dt\int_{\Sigma}{\rm tr}\left(g^{-1}(t){dg(t)\over dt}\
F^n[\gamma(t)]\right)\,,
\eqno(2.34)
$$
which is just $i$ times the expression for the integral (2.16), if written
in the (time dependent) gauge where $A_0=0$. The integration thus leads to
$i$ times the right hand side of (2.23), where $w[g]$ is now the integer
in $\pi_1(\Gs)$ represented by the loop along which we just integrated.
For $-i\Omega$ to be a $U(1)$ connection, the result of the integration
must be $2\pi$ times an integer (see e.g. Ref. 1) which again leads to
the condition of $f$ being even.
The integer is then known as the Chern-class of the $U(1)$
bundle, which here represents an element in $H^2(\Qs,R)$, the second
DeRahm cohomology group.

\beginsection  Appendix

We wish to calculate $\delta W[A]$. For this we calculate the variation
$\delta W[A]$ at a particular point $A$ and obtain the function $W[A]$
in a simply connected neighbourhood by integration. For simplicity we
chose as evaluation point $A$
a static ($\partial_0A=0$) pure magnetic field ($F_{0k}=0$) in the gauge
$A_0=0$. $\{T_1,\dots ,T_k\}$, $k=dim\,SU(N)$, denote Hermitean matrices
such that the matrices $iT_p$ form a basis for the Lie algebra of $SU(N)$.
Flavour and spinor indices on the fermionic fields will not be displayed.
We then have
$$\eqalignno{
{\delta\over \delta A^p_0}\Big\vert_{A_0=0}\,
W[A]=&
       {\delta\over \delta A^p_0}\Big\vert_{A_0=0}\,
       \ln\left[\int d\bar{\psi}d\psi\quad{\rm e}^{\int \bar{\psi}
       (i\Ds-m)\psi}\right]&\cr
    =& {\delta\over \delta A^p_0}\Big\vert_{A_0=0}\ln{\rm DET}(i\Ds-m)&\cr
    =& {\delta\over \delta A^p_0}\Big\vert_{A_0=0}\ln{\rm DET}
       (i\Dts-m+i\gamma^0A_0)(-i\Dts-m)&\cr
    =& {\delta\over \delta A^p_0}\Big |_{A_0=0}\ln{\rm  DET}
        \left[\Dts^2+m^2-i\gamma^0 A_0(i\Dts +m) \right]
     &(A.1)\cr}
$$
where $\Dts:=\Ds|_{A_0=0}=\gamma^0\partial_0+\Ds_{2n}$. We now regard
the third term in the last expression of (A.1) as a perturbation to
$X=\Dts^2+m^2$. The determinant itself is defined via $\zeta$-function
regularisation:
$$\eqalignno{
\zeta_X(s):&=\sum_n\lambda_n^{-s}\cr
        {\rm det}\,X:&=\exp(-\zeta'_X(0))\,. &(A.2)\cr}
$$
Here $\lambda_n$ is the n'th eigenvalue of the operator $X$.
If $|\phi_n\rangle$ are the corresponding eigenvectors, we have
$$
\delta\lambda_n=\langle\phi_n|\,\delta(\Dts^2+m^2)\,|\phi_n\rangle
=\langle\phi_n|\,-i\gamma^0\delta A_0(i\Dts +m)\,|\phi_n\rangle\,.
\eqno(A.3)
$$
This gives for the variation $\delta\ln\hbox{\rm DET}$ on the right
hand  hand side of (A.1)
$$\eqalignno{
& -{d\over ds}\Big\vert_{s=0}\delta\zeta_{X(A_0)}
=  {d\over ds}\Big\vert_{s=0}\sum_n s\
\delta\lambda_n\,\lambda_n^{-(s+1)}
&\cr
= & f\,{d\over ds}\Big\vert_{s=0}\quad {\rm TR}_{\phi\lambda\sigma}
\left[\delta A_0^pT_p\gamma^0\,s(i\Dts+ m)(\Dts^2+m^2)^{-(s+1)}\right]\,
&(A.4)\cr}
$$
where ${\rm TR}_{\phi\lambda\sigma}$ denotes the trace operation over
the  space-time functions ($\phi$), the Lie algebra ($\lambda$) and
the spinor space ($\sigma$).
The factor $f$ results from having already taken the trace over
the $f$-dimensional flavour space. Negative powers of positive
operators are defined via
$$
X^{-(s+1)}:={1\over \Gamma(s+1)}\int_0^{\infty}dt\,t^s\exp(-tX)\,,
\eqno(A.5)
$$
so that after some rearrangements we obtain for the expression in (A.4)

$$\eqalign{
f\,{d\over ds}\Big |_{s=0}\,{s\over \Gamma(s+1)}\,\int_{R^{2n+1}}
d^{2n+1}x\quad {\rm TR}_{\lambda\sigma}
&
  \left[
  T_p\gamma^0\delta A_0^p\;\int_{R^{2n+1}}
  {d^{2n+1}k\over (2\pi)^{2n+1}}\right.\;
  {\rm e}^{-ik_{\mu}x^{\mu}}                      \cr
  \times(i\gamma^0\partial_0+i\Ds_{2n}+m)
& \left.\int_0^{\infty}dt\,t^s\,
{\rm e}^{-t(-\partial_0^2+\Ds^2_{2n}+m^2)}
\,{\rm e}^{ik_{\mu}x^{\mu}}\right]                 \cr}
\eqno(A.6)
$$
where we explicitly expressed the trace over the space-time functions
in a plane wave basis $\exp (ik_{\mu}x^{\mu})$. We write $k_0$ for the
time-component and $\vec k$ for the collection of space-components $k_i$
of $k_{\mu}$. Now, the first two terms in the round bracket do not
contribute since
$$
i\partial_0{\rm e}^{-t(\cdots)}{\rm e}^{ikz}={\rm e}^{ikz} i(\partial_0 +
ik_0){\rm e}^{-t(\cdots)}
\eqno{(A.7)}
$$
which vanishes upon $k_0$ integration and our
staticity requirement, and
$$
{\rm TR}_{\sigma}(\gamma^0\Ds_{2n}{{\rm e}^{-t(\cdots)}})=
-{\rm TR}_{\sigma}(\Ds_{2n}{\rm e}^{-t(\cdots)}\gamma^0)=0\,,
\eqno{(A.8)}
$$
since there are an even number of spatial $\gamma^i$'s
in $(\cdots)$. Further, we have
$$
\Ds^2_{2n}=D_kD^k+{1\over 2}\gamma^k\gamma^l F_{kl} \,,
\eqno{(A.9)}
$$
and can thus write (A.1)
$$\eqalign{
& {\delta\over\delta A_0^p}\Big\vert_{A_0=0}W[A]= \cr
&  fm\,{d\over ds}\Big |_{s=0}\,{s\over \Gamma(s+1)}\,
    {\rm TR}_{\lambda\sigma}
  \left[T_p\gamma^0\int_0^{
  \infty}dt\,t^s\right. \int_{R^{2n}}{d^{2n}\vec k\over (2\pi)^{2n}}
   {\rm e}^{-t{\vec k}^2}\,\cr
&
\times\int_{-\infty}^{\infty}{dk^0\over 2\pi}{\rm e}^{-t(k_0^2+m^2)}
\left.\exp(-t(D_kD^k+2ik^lD_l+{1\over 2}\gamma^l\gamma^kF_{kl}))
\right]\,.\cr}
\eqno(A.10)
$$

In the 2n-dimensional spatial space, $\gamma^{2n+1}$ (``gamma five'')
is given by $ \gamma^{2n+1}=i^n\gamma^1\cdots\gamma^{2n}$ which is
Hermitean and squares to one. Also,
${\rm TR}_{\sigma}(\gamma^{2n+1}\gamma^{i_1}\cdots\gamma^{i_{2n}})
=(-i)^n2^n\varepsilon^{i_1\dots i_{2n}}$. So if we choose
$\gamma^0=-i\gamma^{2n+1}$, we obtain
$$
{\rm TR}_{\sigma}\,\left(\gamma^0\gamma^{i_1}\cdots\gamma^{i_{2n}}\right)=
(-i)^{n+1}2^n\varepsilon^{i_1\dots i_{2n}}\,,
\eqno{(A.11)}
$$
and have for the first non-vanishing contribution from the exponential
in (A.10)
$$\eqalignno{
&(-i)^{n+1}{t^n\over 2^n n!}{\rm TR}_{\lambda\sigma}
\left[T_p\gamma^0\gamma^{i_1}
\cdots\gamma^{ 1_{2n}}\,F_{i_1i_2}\cdots F_{i_{2n-1}i_{2n}}\right]
\cr &\cr
=&-i\,{t^n\over n!2^n}\,2^n\,\varepsilon^{i_1\dots i_{2n}}\,
{\rm TR}_{\lambda}
(T_p\;iF_{i_1i_2}\cdots iF_{i_{2n-1}i_{2n}})\cr &\cr
=&-it^n2^n \left[* {\rm TR}_{\lambda}\left\{T_p\,\exp(iF)\right\}]\right]^0
 &(A.12)\cr}
$$
where from the first to the second line we have performed the trace
over the $2^n$-dimensional spinor space. The last expression is meant
to be the $0$-component of the 1-form in curly brackets, where $*$
denotes the Hodge-duality operator with respect to the
$2n+1$-dimensional metric $\delta_{\mu\nu}$. Performing the
$dk$-integration yields a factor of $(4\pi t)^{-n}$ so that we can write
expression (A.12) as follows:
$$\eqalignno{
&-ifm\,{d\over ds}\Big |_{s=0}\,{s\over \Gamma(s+1)}\,
\int^{\infty}_{-\infty}{dk^0
\over2\pi}\int^{\infty}_0dt\,t^s{\rm e}^{-t(k_0^2+m^2)}\;
\left[*{\rm TR}_{\lambda}\left\{T_p\,\exp\left({i\over 2\pi}F
\right)\right\}\right]^0
\cr
&=-ifm\,
\left[*{\rm TR}_{\lambda}\left\{T_p\,\exp\left({i\over 2\pi}F
\right)\right\}\right]^0
\, \int_{-\infty}^{\infty}{dk_0\over 2\pi(k_0^2+m^2)}\cr
&=-i{f\over 2}\,
\left[*{\rm TR}_{\lambda}\left\{T_p\,\exp\left({i\over 2\pi}F
\right)\right\}\right]^0
\,.
&(A.13)\cr}
$$
Although we had selected the $0$-th component to arrive at this
expression, relativistic covariance tells us that the corresponding
relations hold for any component (this is also apparent from the
derivation, where any other component could have been preferred).

If we now include higher powers $t^{n+r}$ from the exponential in
(A.10) the $k$-integration deletes again $n$ of them so that we are
left with an integral of the form
$$
{s\over \Gamma(s+1)}\int_0^{\infty}dt\,t^{r+s}\,{\rm e}^{-t(k_0^2+m^2)}=
{\Gamma(s+1+r) \over \Gamma(s+1)}{s\over (k_0^2+m^2)^{s+r+1}}
\eqno{(A.14)}
$$
which, when acted upon by ${d\over ds}\big |_{s=0}$, gives
a term $\propto(k_0^2+m^2)^{-r-1}$,
and after $k^0$-integration a term $\propto m^{-2r-1}$.

So, finally, writing tr for ${\rm TR}_{\lambda}$, we arrive at the
compact formula
$$
\delta W[A]=-{f\over 2}\,*\,
{\rm tr}\left\{\delta A\exp\left({i\over 2\pi}F\right)\right\}+
{\rm terms} \propto {1\over m}\,.
\eqno{(A.15)}
$$

\vfill\eject

\beginsection References

{\parskip=0.1truecm

\item{1}  N. Woodhouse, {\it Geometric Quantization}.
            Claredon Press, Oxford (1980).

\item{2}  L. Alvarez-Gaum\'e, and P. Nelson, {\it Comm. Math. Phys.}
          {\bf 99}, 103 (1985)

\item{3}  N. Narasimhan, and T. Ramadas, {\it Com. Math. Phys.}
            {\bf 67}, 121 (1979).

\item{4}  J. Mickelsson, {\it Comm. Math. Phys.} {\bf 97}, 361 (1985).

\item{5}  I. Singer, ``The geometry of the orbit space for non-abelian gauge
            theories'', lectur delivered at the conference on
            {\it Perspectives in Modern Field Theories''}, Stockholm, (1980).

\item{6}  B. Babelon, and C. Viallet, {\it Comm. Math. Phys.} {\bf 81},
            515 (1981).

\item{7}  M. Atiyah, and I. Singer, {\it Proc. Nat. Acad. Sci. USA}
            {\bf 81}, 2597 (1984).

\item{8}  Y.S. Wu, {\it Phys. Lett. B} {\bf 153}, 70 (1985).

\item{9}  R. Bott, and R. Seeley, {\it Comm. Math. Phys.} {\bf 62},
            235 (1978).

}

\end